\documentstyle[11pt,aasms]{article}
\lefthead{??}
\righthead{}

\def\lta{{\>\rlap{\raise2pt\hbox{$<$}}\lower3pt\hbox{$\sim$}\>}}
\def\gta{{\>\rlap{\raise2pt\hbox{$>$}}\lower3pt\hbox{$\sim$}\>}}

\lefthead{??}
\righthead{}
 
\begin{document}

\title {The Metallicity of $0.5<z<1$ Field Galaxies}

\author{C. Marcella Carollo$^\dagger$\affil{Columbia University, 
Department of Astronomy, New York, NY 10027}}

\author{Simon J. Lilly$^\dagger$\affil{University of Toronto,
Department of Astronomy, Toronto, ON M5S 3H8 Canada\\ Herzberg
Institute of Astrophysics, Victoria V9E 2E7}}

\bigskip

$^\dagger$ Visiting Astronomer at the Canada-France-Hawaii Telescope
which is operated by the National Research Council of Canada, the
Centre National de la Recherche Scientifique and the University of
Hawaii.

\begin{abstract}

We have measured the emission line ratios in a sample of 34 CFRS
star-forming galaxies with redshifts between $0.5 < z < 1.0$, and
computed their metallicities by means of the empirically-calibrated
$R_{23}$ metallicity estimator introduced by Pagel et al.\ (1979).
The current analysis concentrates on the 15 galaxies with $L_{H\beta}
> 1.2 \times 10^{41}$ erg s$^{-1}$.  Although our results can only be
regarded as preliminary until near-IR spectroscopy of H$\alpha$ and
[NII]6583 are available, the metallicities of these galaxies appear to
be remarkably similar to those of local galaxies selected in the same
way, and there appears to have been little change in the
relationship between metallicity and line- and continuum-luminosity
from $z \sim 1$ to today.  At this stage our results do not support
the idea that these galaxies, known to be generally small and with
late-type morphologies, are dwarf galaxies brightened by large bursts
of star-formation, as had been suggested from previous
studies. Rather, our findings are more consistent with a picture in
which these systems are the progenitors of today's massive metal-rich
galaxies.

\noindent

\end{abstract}

{\it subject headings}: galaxies: formation - galaxies: evolution -
galaxies: metallicity - galaxies: emission lines

\section{Introduction}

There is now clear evidence from the Canada-France Redshift Survey
(CFRS; Lilly et al.\ 1995a), and other
similar surveys, for evolutionary changes in the field galaxy
population over the redshift interval $0 < z < 1$ manifested as an
increased number of blue galaxies, with rest-frame $(U-V)_{AB} \lta 1.4$,
and moderate luminosities, $L
\sim L^*$.  The redshift regime $0.5 < z < 1.0$ appears to be an
important epoch in the history of the galaxy population.  Evolutionary
effects are clearly seen in the galaxy population, and the apparent
change of behaviour in the ultraviolet luminosity density of the
Universe as a whole around $z \sim 1$ (Lilly et al.\ 1996; Madau et
al.\ 1996; Steidel et al.\ 1999) may indicate that this epoch
represents a transition between the ``high redshift'' Universe at $z >
1$ and that seen today.  While the large galaxies appear to be absent
at $z \sim 2$,
large spiral galaxies and early-type galaxies are
present at $z \sim 0.8$ in numbers comparable to those seen locally,
co-existing with increased numbers of comparably bright but smaller
irregular galaxies (Brinchmann et al.\ 1998, hereafter B98; Lilly et
al.\ 1998; hereafter L98). There is morphological evidence for
substantially elevated levels of merger activity at this epoch (Le
Fevre et al.\ 2000).
 
Much of the work on the $z\sim1$ field galaxy population has been to
date based on gross statistical measures such as the bivariate
color-luminosity function (Lilly et al.\ 1995c; Heyl et al. 1997; Lin
et al.\ 1999).  The HST-based studies of the morphologies and sizes of
field galaxies in this redshift regime (e.g. B98; L98), and of the
internal kinematics of various subsets of these galaxies (Vogt et al.\
1997; Guzman et al.\ 1997; Mallen-Ornelas et al.\ 1999) have suggested
that the most massive galaxies are evolving relatively slowly (Vogt et
al.\ 1997; L98) while the ``excess'' galaxies in the luminosity
function (with $L \sim L^*$ and blue colors) have generally the
irregular morphologies (B98), small sizes (3-5 $h_{50}^{-1}$ Mpc, L98)
and low velocity dispersions $\sigma < 100$ kms$^{-1}$ (Guzman et al.\
1997; Mallen-Ornelas et al.\ 1999) that are usually associated in the
local Universe with galaxies 2-3 magnitudes further down the
luminosity function. This suggests at first sight quite a strong
luminosity brightening in late-type low mass galaxies.

There are still, however, major gaps in our knowledge of galaxies in the
crucial $0.5 < z < 1.0$ redshift regime which make it possible that
this interpretation of the morphologies and the kinematic data is
incorrect.  Not least, the possibility that the small-blue-irregular
galaxies are the cores of more massive galaxies ---i.e., the
``down-sizing'' scenario of Cowie et al.\ (1996)--- cannot be ruled
out at this stage. For example, the $K$-band luminosities of typical
blue galaxies at $z \sim 0.8$ are comparable to those of today's
massive galaxies, leading to suggestions that at least some of these
blue objects may be indeed the progenitors of massive systems.

There has hitherto been almost no systematic study in the $0.5<z<1$
regime of the physical diagnostics that are familiar from studies of
the local Universe. For local galaxies, combinations of strong
emission lines are routinely used to determine or constrain the nature
of the ionizing radiation (Veilleux \& Osterbrock 1987), the amount of
reddening, and the metallicity of the interstellar medium (ISM; Pagel
et al.\ 1979; Kennicutt 1998; Stiavelli 1998; Kobulnicky et al.\
1999).  Although a rigorous determination of the ISM metallicity
requires knowledge of electron temperature derivable only from
intrinsically weak lines, diagnostic line ratios based on stronger
lines have been empirically calibrated.  In particular, the $R_{23}$
parameter, $R_{23} = ([OII]3727 + [OIII]4959+5007)/H\beta$ (Pagel et
al.\ 1979), has been empirically calibrated against metallicity, with
an intrinsic scatter of only 0.2 dex. It is a weak function of the
ionization ratio [OIII]4959+5007/[OII]3727.  A reversal in $R_{23}$
occurs at $Z \sim 0.3 Z_{\odot}$ due to cooling effects, so a low- and
a high-metallicity solution are associated with most values of
$R_{23}$.  This degeneracy can however be broken using the
[OIII]5007/[NII]6584 ratio (Kobulnicky et al.\ 1999).  Based on data
on relatively low redshift galaxies, Kobulnicky et al.\ have discussed
in detail the potential accuracy that could be obtained in using
$R_{23}$ to measure metallicities of unresolved galaxies at high
redshifts.  These authors examined theoretically the effects of dust
reddening, of $H\beta$ absorption, of spatial averaging over extended
galaxies with abundance gradients, and the possible effects of diffuse
interstellar gas, and provided prescriptions for dealing with these
effects.

Determining the metallicity of the ISM of distant starforming field
galaxies is of particular importance, both as a general indicator of
the evolutionary state of these systems, and as a constraint on their
possible present-day descendants. Some work has been done to determine
the metal content of galaxies at redshifts of about $0.1 < z< 0.5$
(Kobulnicky \& Zaritsky 1999); no information has however been
available about the ISM metallicities of galaxies in the $0.5 < z <
1.0$ redshift interval.  We have therefore begun a program of
systematic emission line spectroscopy of CFRS galaxies to determine
the ISM metallicity of $0.5 < z < 1.0$ field galaxies, and to study
how the metal content in these systems correlates with galaxy
luminosity, star-formation rate, structure and morphology.  Our
program uses the $R_{23}$ metallicity estimator and therefore requires
spectroscopy of the [OII]3727, [OIII]4959,5007 and H$\beta$ lines,
which in this redshift range are shifted into the 5000 $\AA$ - 1
$\mu$m wavelength region.  Supplementary spectroscopy of H$\alpha$,
[NII]6584 and [SII]6717,6731 lines, which are shifted into the
near-infrared J-band, allows determination of the reddening, isolation
of active galactic nuclei, and the breaking of the $R_{23}$-degeneracy
with metallicity.

In this Letter we report the first results of our program which are
based on deep multi-object spectrophotometry over the $0.5 < \lambda <
1.0 \mu$m range.  Infrared spectrophotometry has already been acquired
for one object, and will be obtained in due course for the remaining
galaxies, but the optical data on its own already provides a set of
homogeneous data that can immediately be compared at a
phenomenological level with the equivalent local sample (see e.g.
Jansen et al.\ 2000). Throughout this paper we use $H_o=50$ Mpc/Km/s
and $q_o=0.5$, and refer to solar metallicity $Z_\odot$ as $12 +
Log(O/H) = 8.9$.

\section{Sample Selection, Observations and Data Reduction}

The observations were carried out on the nights 5-7 March 2000 at
the 3.6m Canada-France-Hawaii Telescope (CFHT) using the MOS
spectrograph (Le Fevre et al.\ 1994) with the STIS-2 CCD detector with
21 $\mu$m pixels.  A red-blazed grating with 300 l/mm and a
short-wavelength cutoff filter yielded spectra between $5000 \AA$ and
1 $\mu$m and about 20 objects were observed on each multi-slit mask
with slits 20$''$ long and 1.3$''$  wide, giving a spectral
resolution of $R \sim 600$. One mask was observed in each of the
CFRS-10 and CFRS-14 fields (Le Fevre et al.\ 1995; Lilly et al.\ 1995b).

Spectroscopy at $\lambda \ge 7600 \AA$ is challenging on account of
the OH forest.  The most important factors that hamper accurate
spectroscopic measurements are sky-subtraction and fringe removal.
The use of relatively long slitlets ($\sim 20''$) and nine 2700s
exposures allowed us to offset the target galaxies along the slit at
each integration, and thus to largely eliminate defects in the
sky-subtraction that are fixed relative to the chip (e.g. those
arising from chip fringing, imperfections in the slit profile, and
distortions introduced by the spectrograph camera).  The spectra were
reduced with standard IRAF routines and calibrated via observations of
two spectrophotometric standards.  The spectra were corrected for
(small) airmass effects but not for Galactic reddening, since this is
small ($E_{B-V} < 0.03$, Burstein \& Heiles 1984). Comparisons of the
spectra of standard stars observed through different slits in the
masks indicated that the relative spectrophotometric calibration over
the wavelength range of interest was about 10\% (r.m.s.).  However,
the uncertainties in the relative line flux measurements were usually
dominated by systematic uncertainties in establishing the local
continuum, and these were conservatively estimated by exploring rather
extreme possibilities.  The errors in the line ratios were derived
from adding in quadrature these two sources of uncertainty.  In
several cases, the strength of the [OIII]4959 line was only poorly
estimated and in these cases the strength of this line was assumed to
be 2.85 times that of [OIII]5007.  In a few cases, the H$\beta$ and/or
[OIII]5007 lines were not convincingly seen in the spectra; for these
galaxies, conservative upper limits were estimated through comparison
with nearby features in the spectra that were known to be unreal.

The 34 target galaxies were selected from the CFRS to have, in
addition to the original $I_{AB} < 22.5$ photometric selection, an
[OII]3727 flux in excess of $7 \times 10^{-17}$ erg s$^{-1}$cm$^{-2}$
in order to ensure strong enough lines for a measurement of the
$R_{23}$ parameter.  The observational noise in the $R_{23}$
measurement is usually dominated by the uncertainty in the strength of
H$\beta$ which appears in the denominator of $R_{23}$.  The H$\beta$
luminosity is likely to be closely related to the star-formation rate
and represents an astrophysically useful reference to galaxies in the
local Universe.  Unfortunately, H$\beta$ measurements were not
available beforehand for most of the target galaxies because of the
long wavelength cut-off (at about 8400 $\AA$) of the original CFRS
spectra. A close approximation to an $H\beta$-luminosity selected
sample can only be constructed a posteriori.  The analysis presented
in this paper is based on the 15 objects ---about 50\% of the original
[OII]3727 selected sample observed during the run--- with H$\beta$
luminosities above $L_{H\beta} > 1.2 \times 10^{41}$ erg s$^{-1}$.  It
should be noted that any objects missing from such a sample would have
low [OII]3727/H$\beta$ ratios, and thus generally low $R_{23}$ values
which generally correspond to high metallicities. This
high-$L_{H\beta}$ selection well samples the `blue' CFRS galaxy
population with restframe $(U-V)_{AB} \lta 1.4$ (Table 1).  Two of the
highest redshift galaxies, at redshifts $z=0.8718$ and $0.9203$,
respectively, have easily detectable [NeIII]3869. On the
$Log([NeIII]3869/H\beta)$ versus $Log([OII]3727/H\beta)$ plane of Rola
et al.\ (1997), these systems (unlike the other 13) occupy locations
typical of the local AGN population.  Therefore, although they are
formally part of our high-$L_{H\beta}$ sample, these probable-AGN will
be distinguished in our comparison with the local galaxies.  The
remaining 13 high-$L_{H\beta}$ objects span the redshift range $0.6 <
z < 1$ with a median redshift $z \sim 0.783$.

Three high redshift galaxies in the original sample of 34 had upper
limits to their H$\beta$ luminosities which were above our $L_{H\beta}
= 1.2 \times 10^{41}$ erg s$^{-1}$ threshold, possibly putting these
systems into the high-$L_{H\beta}$ sub-sample that is discussed in
this paper.  The line ratios for these galaxies are necessarily
uncertain; however, we have included these systems in the figures for
completeness, and assigned to them the range of line ratios that they
would have if their H$\beta$ luminosities were indeed above our
threshold. We have however identified these three objects with
different symbols, as a reminder that they may actually not belong to
the high-$L_{H\beta}$ sub-sample.

We report in Table 1 the relevant parameters for the 15 $L_{H\beta}
> 1.2 \times 10^{41}$ erg s$^{-1}$ galaxies and for the three galaxies
with upper limits on $L_{H\beta}$ above this threshold.

\section{Results and Discussion}

Figures 1a and 1b show the [OIII]5007/[OII]3727 versus $R_{23}$
relation for the local field sample of Jansen et al.\ (2000) and the
high-$L_{H\beta}$ CFRS galaxies of our sample, respectively.  The
galaxies of the Jansen et al.\ sample were selected from the CfA
redshift catalog (Huchra et al.\ 1983) to span a large range in
absolute B magnitude (from -14 to -22), while sampling fairly the
changing mix of morphological types as a function of luminosity.  In
the figures, the solid and dashed lines are lines of constant
metallicity.  From left to right, the dashed lines indicate increasing
metallicities from 0.02$Z_\odot$ to 0.2$Z_\odot$. At $Z \sim
0.3Z_\odot$, the reversal of $R_{23}$ occurs, and the curves of
constant metallicity then follow the solid-line sequence in which $Z$
rises up to about 3$Z_\odot$ from right back to left. The $Z =
Z_\odot$ curve is highlighted with a thicker linewidth.  In order to
appropriately compare the local and the high--$z$ samples, the local
Jansen et al.\ galaxies which have L$_{H\beta} > 1.2\times10^{41}$
ergs s$^{-1}$ are identified with large circles; local galaxies with
smaller $H\beta$ luminosities are represented by small
circles. Furthermore, in Fig 1a, filled symbols represent objects
which have a flux ratio $[OIII]5007/[NII]6584 > 2$ and thus
metallicities $Z \lta 0.5 Z_\odot$, and empty symbols indicate objects
with higher metallicities, as indicated by a flux ratio
$[OIII]5007/[NII]6584 < 2$ (see Edmunds \& Pagel 1984). In Figure 1b,
the high--$z$ galaxies are represented by the filled squares, with the
exception of the two probable-AGN (identified by the asterisks), and
of the three objects which may or may not be in the sample on account
of their H$\beta$ upper limits (empty squares).  The fiducial
dotted-line box in both panels encompasses the bulk of the Jansen's
galaxies.  The arrows on the right-side of the figures represent the
direction in which the points of the diagram would shift due to
reddening by dust (as described by Cardelli et al.\ 1989); the length
of the arrows refer to an $E(B-V)=0.3$ magnitudes at $z=0.7$. The
effects of reddening are a function of the [OIII]5007/[OII]3727 ratio.

Several things are apparent in Figures 1a and 1b.  First, the high
redshift galaxies fall in the $R_{23}$ versus [OIII]/[OII] plane in
locations that are occupied by galaxies in the local Jansen et al.\
(2000) sample.  Furthermore, once the AGNs are  excluded, the
high redshift sample selected to have $L_{H\beta} > 1.2 \times
10^{41}$ erg s$^{-1}$ occupies the same restricted location in the
$R_{23}$ versus [OIII]/[OII] plane  as do the objects in the local
Universe with similarly high $H\beta$ luminosities.  At both epochs,
galaxies selected to have the same H$\beta$ luminosities exhibit the
same range of $R_{23}$ and [OIII]/[OII]. (We note that there may
possibly be a small displacement of the high redshift galaxies towards
the upper edge of the dotted box in Figure 1a. This could conceivably
be due to the effects of higher dust extinction at high redshifts; it
would be premature to claim this  at this stage.)

It is also apparent on Figure 1b that once the two probable-AGNs are again
excluded, there are no high-$L_{H\beta}$ objects at high redshifts (at
least in this still small sample) that have $R_{23} \sim 7$, the value
that is non-degeneratively associated with intermediate metallicities,
i.e. $Z \sim 0.3 Z_{\odot}$. The $Z$-degenerate $R_{23}$ values that
are measured for the high-$z$, high-$L_{H\beta}$ galaxies indicate
either rather low ($\lta 0.1Z_\odot$) or rather high ($\sim Z_\odot$)
metallicities for these systems. This is again similar to what is
observed in the local sample: high-$L_{H\beta}$ galaxies avoid the
intermediate metallicity regime both at the present and at the $z\sim
1$ epoch.

The similarity between the high and low redshift samples is further
illustrated in Figure 2, which shows the relationship between $R_{23}$
and continuum luminosity M$_B$. There is a rather startling similarity
between the high-$z$ and the local galaxies on this diagram.  There is
little evidence for an evolutionary change in the relationship between
metallicity (as estimated from $R_{23}$) and the line and continuum
luminosities for high-$L_{H\beta}$ field galaxies between $z \sim 0$
and $z \sim 0.8$. This extends to higher redshifts the findings of
Kobulnicky \& Zaritsky (1999) which was based on a similar sample at
$0.1 < z < 0.5$.

For one of the non-AGN galaxies, CFRS-14.0393, we have already obtained a
Keck spectrum in the J-band, and analysed it to derive the intensity
of the [NII]6584 emission line.  We will report the details on this
analysis elsewhere. The important fact, relevant for the current
discussion, is that the [OIII]5007/[NII]6584 ratio that this J-band
spectrum has allowed us to measure for this ($z=0.6035$) object
undoubtedly places it on the high-metallicity branch of the $R_{23}$
parameter, with a metallicity quite close to the solar value. This one
case argues in favour of a high, i.e. about solar metallicity for at
least some objects in our $0.5<z<1$ high-$L_{H\beta}$ sample. The HST
morphology of this object is that of a regular two-armed spiral
(Schade et al.\ 1995), and so it is possibly not surprising that this
galaxy has a high metallicity. It may be that our high-$L_{H\beta}$
selection favours large well-formed galaxies relative to the general
blue CFRS population. However, there is no indication that this is the
case from the HST morphology of three additional objects for which the
HST data are available (CFRS-10.1213, CFRS-14.0972 and CFRS-14.1258).

Of course, until we obtain the infrared spectroscopy for the entire
sample, the $R_{23}$ degeneracy with $Z$ does not allow us to prove on
what branch ---i.e., the low- or the high-metallicity one--- each
individual high-$z$ galaxy lies, and the possibility remains that some
objects have indeed metallicities $Z\lta 0.1Z_\odot$.  However, at
this stage, the absence of any galaxies with $Z \sim 0.3 Z_{\odot}$
makes this possibility rather contrived, since it would imply a
bimodal distribution of metallicities with a ``gap'' around such
intermediate values of $Z$.  Therefore, although confirmation must
await the infrared spectroscopy, the best working hypothesis seems the
one where {\it all} of the analysed high-$L_{H\beta}$ high redshift
galaxies have relatively high metallicities, i.e. within 40\% of
solar.

The consequences of these findings are interesting in the context of
the studies of sizes, morphologies and kinematics of high-redshift
galaxies discussed above. It is clear in fact that, at this stage, the
metallicity measurements of $0.5<z<1$ systems do not support the idea
that many of the small and irregular blue L* galaxies that are
responsible for the evolution of the luminosity function back to $z
\sim 0.8$, and which dominate the CFRS at these redshifts (B98; L98)
are low mass (i.e. low metallicity) dwarfs brightened by substantial
luminosity evolution.  In contrast, the metallicity data seem to
suggest the interesting possibility that these small irregular
galaxies are in fact the progenitors of today's massive, metal-rich
galaxies, but seen in an earlier phase of their evolution when they
were already significantly metal-rich but morphologically more
disturbed and smaller. Although about a 1/3 of the CFRS has been
imaged by HST (see Schade et al 1996; B98), the overlap with this
spectroscopic sample is as yet still small.  Testing this idea by
studying the morphologies of these metal-rich systems will have
important consequences for our understanding of the evolutionary path
of massive galaxies.

\bigskip

\acknowledgements We thank Marijn Franx, Jules Halpern, Nino Panagia
and Massimo Stiavelli for helpful discussions.  SJL's research in
Toronto is supported by the Natural Sciences and Engineering Research
Council of Canada and by the Canadian Institute for Advanced Research,
and this support is gratefully acknowledged.

\newpage

\begin{table*}
{\small\begin{center}\begin{tabular}{lllllllll}
\hline\hline
\multicolumn{1}{l}{CFRS-\#} &
\multicolumn{1}{c}{ M$_B$} &
\multicolumn{1}{c}{$z$} &
\multicolumn{1}{c}{$(V-I)_{obs}$} &
\multicolumn{1}{c}{$(U-V)_{rest}$} &
\multicolumn{1}{c}{Log($R_{23}$)} &
\multicolumn{1}{c}{Log([OIII]5007/[OII]3727)} \\
&(mag) && & &\\
\hline 
10.0498$^\dagger$ &     -21.22  &     0.920  &  0.66   &0.41& 1.38 (-0.5)         &           0.20      (0.03)\\   
10.1213 &     -20.94  &     0.815            &  1.27   &0.97& 0.47 (-0.29 / +0.38)&  	    -0.19       (-0.31 / +0.19)  \\ 
10.1925 &     -21.00  &     0.783            &  1.00   &0.62& 0.51       (0.12)   &  	    -0.13       (-0.19 / +0.14)  \\ 
(10.2164) &     -22.64  &     0.859          &  2.95   &2.48& 0.08 (-0.5) 	 &	    -1.20       (-1.45)\\ 
(10.2183) &     -21.87  &     0.910          &  1.47   &1.15& 0.56 (-0.5)         &	    -1.15       (-1.04)\\ 
10.2418 &     -22.00  &     0.796            &  2.39   &2.01& 0.19 (-0.27 / +0.24)& 	    -1.10       (-0.80)\\ 
10.2428$^\dagger$ &     -21.23  &     0.872  &  1.78   &1.30& 0.83 (0.07)         &	     0.16       (0.01)  \\ 
14.0217 &     -20.94  &     0.721            &  1.00   &0.64& 0.56 (-0.09 / +0.10 &  	    -0.17       (0.04)  \\ 
14.0272 &     -21.98  &     0.670            &  1.19   &0.92& 0.09 (-0.25 / +0.29)&  	    -1.25       (-0.76)\\ 
14.0393$^{\dagger\dagger}$ & -21.85&  0.603  &  0.98   &0.72& 0.45 (-0.11 / +0.12)&  	    -0.28       (0.03)  \\ 
14.0438 &     -22.12  &     0.988            &  0.74   &0.51& 0.47       (0.10)   &  	    -0.75       (-0.13 / +0.10)  \\ 
(14.0497) &     -21.22  &     0.800          &  1.10   &0.73& 0.19 (-0.14)  	 &	    -1.30       (-1.44)\\ 
14.0538 &     -21.18  &     0.677            &  0.57   &0.20& 0.70       (0.07)   &          -0.08       (0.02)  \\ 
14.0605 &     -20.69  &     0.837            &  0.40   &0.10& 0.62       (0.09)   &  	     0.22       (0.03)  \\ 
14.0818 &     -22.27  &     0.901            &  1.12   &0.83& 0.47       (0.17)   &  	    -0.22       (-0.31 / +0.18)  \\ 
14.0972 &     -21.37  &     0.810            &  0.82   &0.43& 0.62       (0.05)   &  	     0.02       (0.02)  \\ 
14.1258 &     -20.14  &     0.647            &  0.97   &0.65& 0.54       (0.07)   &  	     0.06       (0.02)  \\ 
14.1386 &     -21.46  &     0.744            &  1.05   &0.69& 0.25 (-0.14 / +0.15)&  	    -0.40       (-0.06 / +0.05)  \\ 
\hline
\end{tabular}\end{center}}
\caption{The 15 galaxies with $L_{H\beta} > 1.2 \times 10^{41}$ erg
s$^{-1}$, and the three additional objects with $L_{H\beta}$ upper
limits above this threshold. These three objects are identified by a
parenthesis in column 1, which lists the CFRS identification number
(Lilly et al.\ 1995b; Le Fevre et al.\ 1995).  The remaining columns
list, respectively, the absolute $B$ magnitude, the redshift, the
observed $V-I$ and the rest-frame $U-V$ colors (AB magnitudes; from
Lilly et al.\ 1995c), the $R_{23}$ parameter and the
[OIII]5007/[OII]3727 ratio. In colum 1, the symbol ``$\dagger$''
identifies the two probable-AGN; the ``${\dagger\dagger}$'' identifies
the one galaxy with the available $J$-band Keck spectrum. In the last
two columns, the positive numbers in parenthesis are the error bars on
the reported measurements (single-valued entries refer to symmetric
errors); negative numbers in parenthesis indicate that the reported
values are limits (lower ones for [OIII]5007/[OII]3727, upper ones for
$R_{23}$).}
\label{tab1}
\end{table*}
\normalsize

\newpage

\begin{figure}[t]
\caption{The [OIII]5007/[OII]3727 versus $R_{23}$ relation for the
local field sample of Jansen et al. (Panel a) and the L$_{H\beta} >
1.2\times 10^{41}$ ergs s$^{-1}$ CFRS galaxies of our sample (Panel
b). Symbols are explained in the text. The $0.5<z<1$ galaxies with L$_{H\beta}
> 1.2\times 10^{41}$ ergs s$^{-1}$ appear to have metallicities as
high as the equivalent high-$L_{H\beta}$ objects in the local
universe.}
\end{figure}

\begin{figure}
\caption{The $R_{23}$ versus absolute $B$ magnitude  relation for
the galaxies in the local universe (circles) and those in the $0.5 < z
< 1$ redshift regime (squares and asterisks).  Symbols are as in
Figures 1a and 1b, and are explained in the text. Above the
L$_{H\beta}>1.2\times10^{41}$ ergs s$^{-1}$ cutoff, the distribution
of points for the local and the high-$z$ galaxy populations is
identical.}
\end{figure}

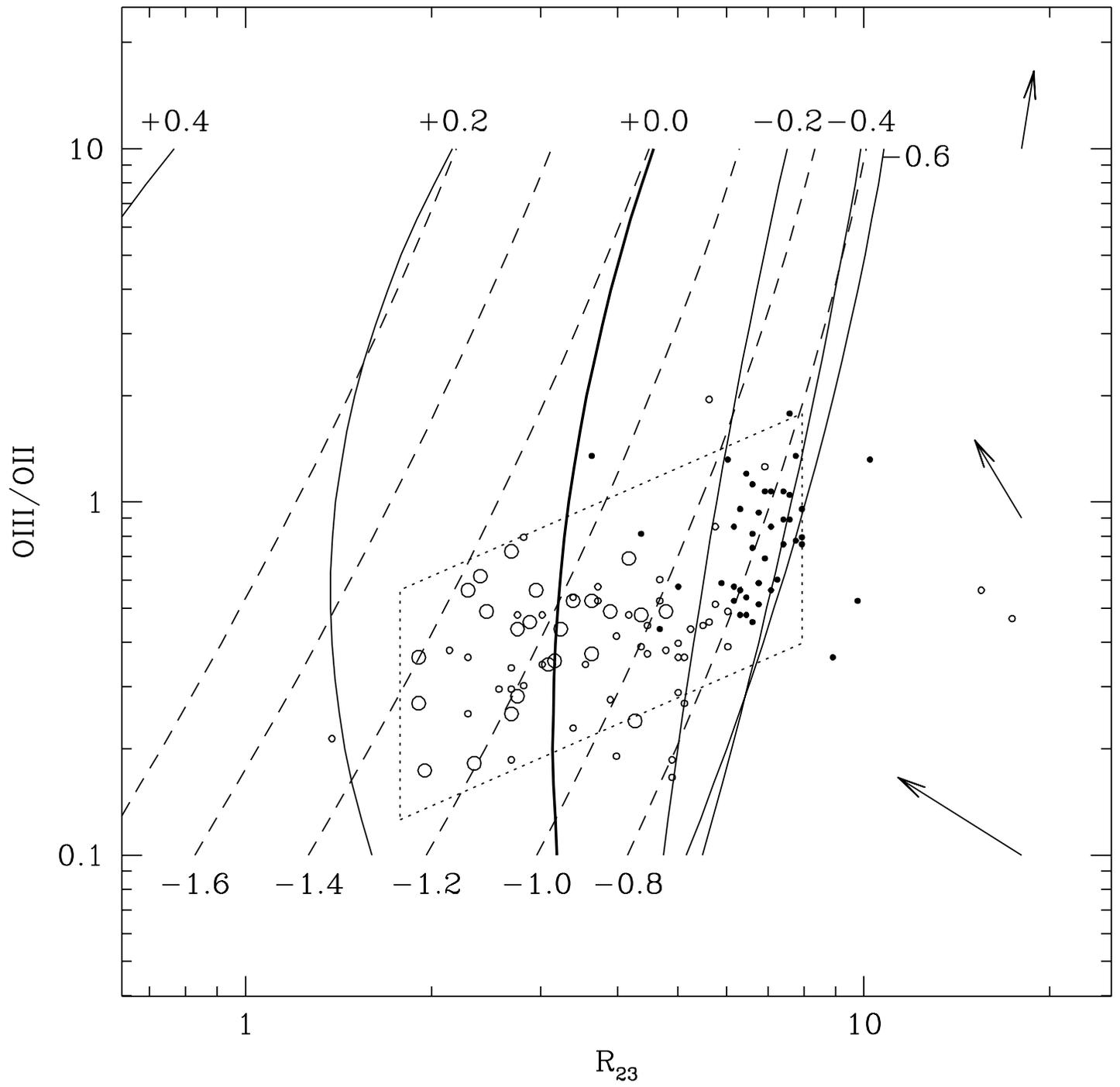
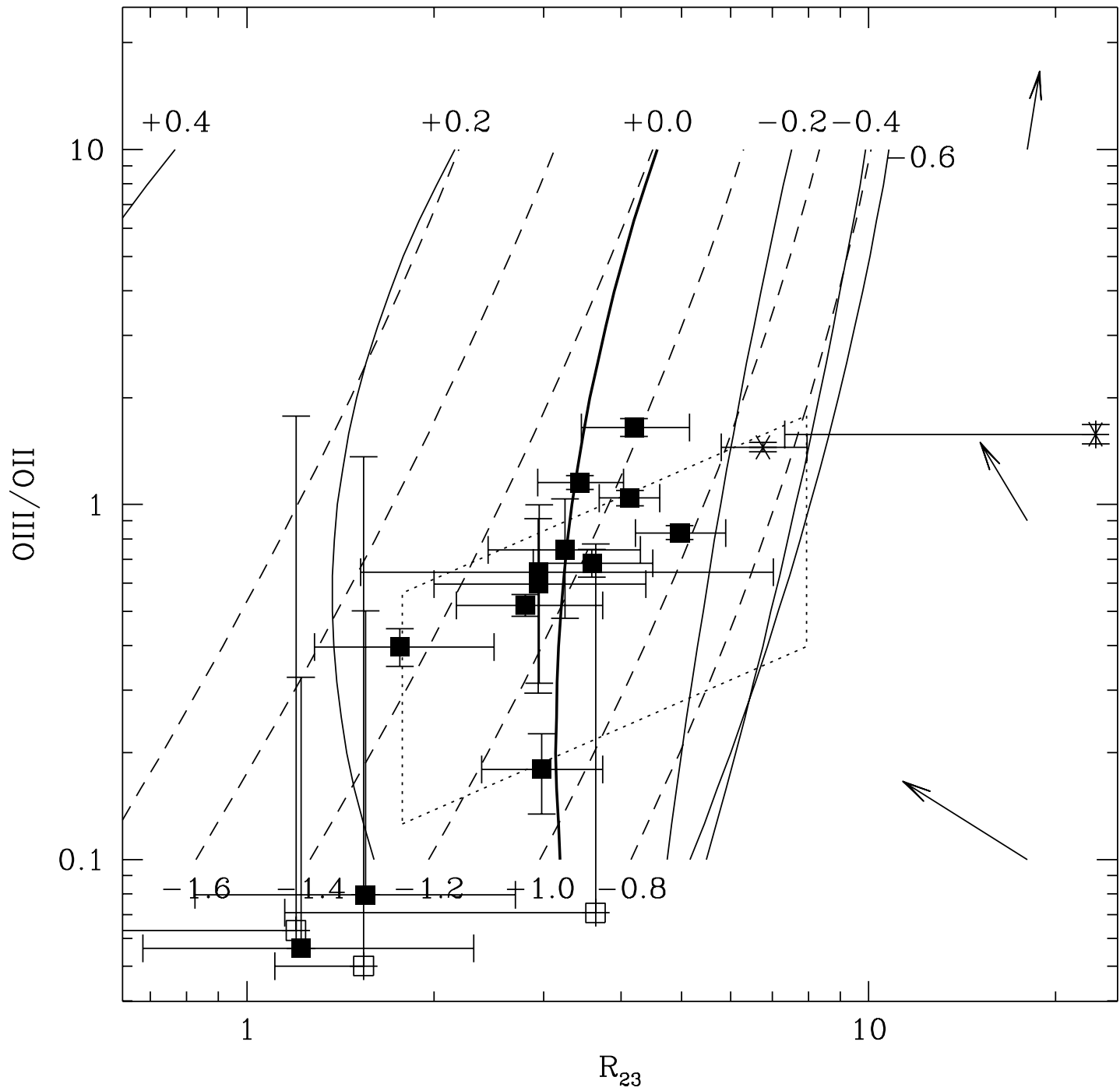
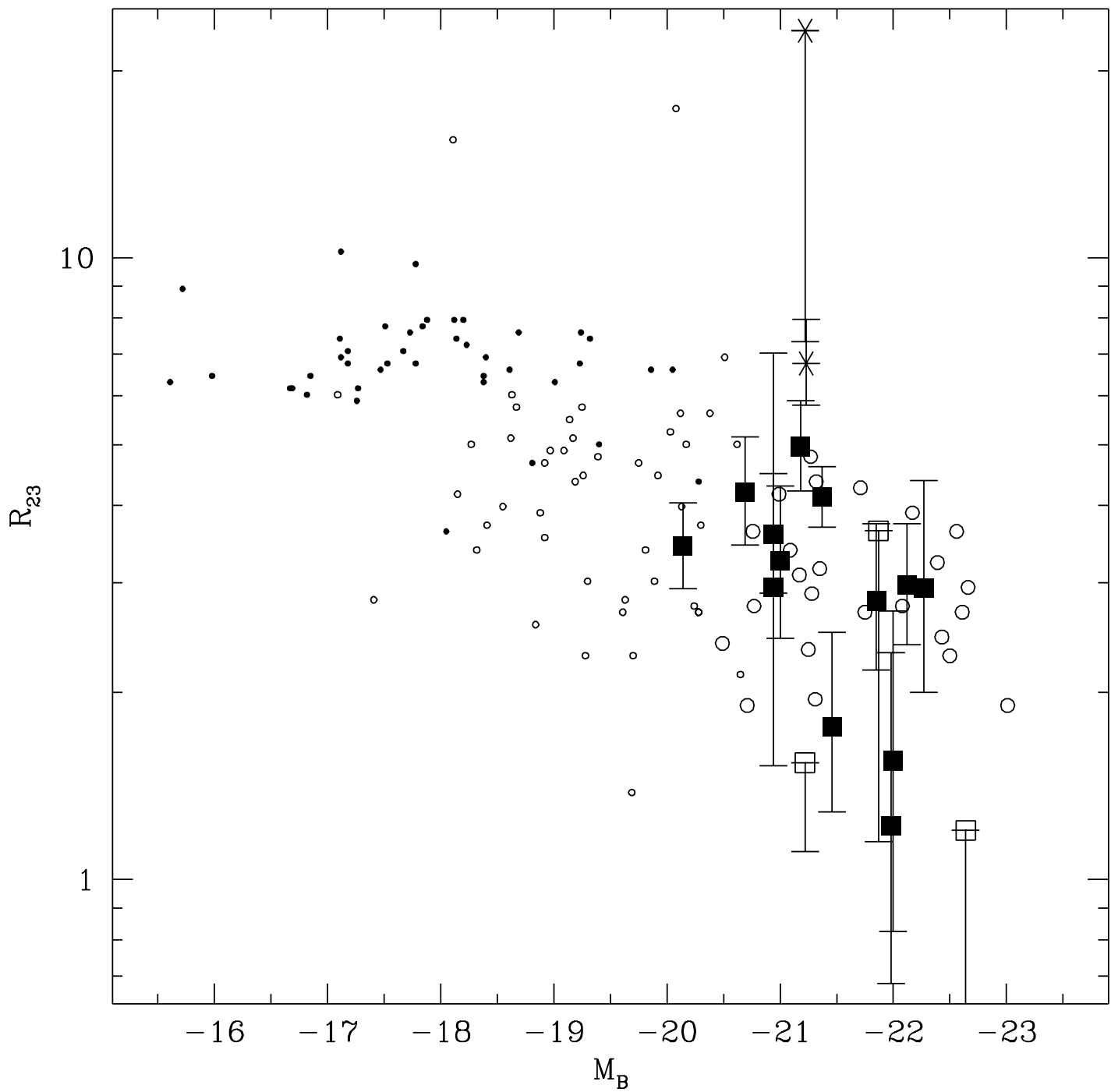
\end{document}